\documentclass{JHEP3} % 10pt is ignored!

\usepackage{epsfig,multicol,bbm}
\usepackage{appendix}

\newcommand\fverb{\setbox\fverbbox=\hbox\bgroup\verb}
\newcommand\fverbdo{\egroup\medskip\noindent%
			\fbox{\unhbox\fverbbox}\ }
\newcommand\fverbit{\egroup\item[\fbox{\unhbox\fverbbox}]}
\newbox\fverbbox

\title{Measuring the Lifetime of Trapped Sleptons Using the General Purpose LHC Detectors}

\author{ James Pinfold\thanks{Communicating author}, \\
	Physics Department, University of Alberta, Edmonton, Alberta T6G 2N6\\
	E-mail: \email{pinfold@phys.ualberta.ca}}
	
\author{ Logan Sibley, \\
	Physics Department, University of Alberta, Edmonton, Alberta T6G 2N6\\
	E-mail: \email{logans@phys.ualberta.ca}}

\received{\today} 		%%
\accepted{\today}		%% These are for published papers.

\abstract{ In supergravity where the gravitino is the  lightest supersymmetric particle (LSP), 
the next-to-lightest supersymmetric particle (NLSP) decays to the gravitino 
with a naturally long lifetime ( $10^{4} - 10^{8}$ s). However, cosmological 
constraints favour charged sleptons with lifetimes below a year 
as the natural NLSP candidate.  For this scenario we report a method to 
accurately determine the slepton lifetime and SUSY cross-section  from observation 
of the decays of sleptons trapped in  the material comprising the main  detector (ATLAS, CMS). 
A measurement of the lifetime  to 5\% is possible after 3 years at nominal luminosity 
and running conditions. This method is sensitive to the cosmologically
preferred stau lifetime of $\sim$ 37 days and does not require the use of ancillary
 trapping  volumes.}


\begin{document} 

%\maketitle  IS IGNORED %%%%%%%%%%%

\section{Introduction}

Most studies of supergravity models assume  the  lightest supersymmetric particle (LSP)
 is a standard model superpartner, such as a slepton or neutralino. However, there are
  theoretically plausible scenarios  where the gravitino is the LSP \cite{GRAVITINO2}
  \cite{GRAVITINO3} \cite{GRAVITINO4} \cite{GRAVITINO5} \cite{GRAVITINO6} \cite{GRAVITINO7}
   \cite{GRAVITINO8} \cite{GRAVITINO9} \cite{GRAVITINO10} \cite{GRAVITINO11} \cite{GRAVITINO12}.
Here the gravitino is a ``superWIMP''  with a  weak scale mass of $\sim$ 100 GeV and 
couplings suppressed by the reduced Planck scale (RPS),  $M_{RPS}  =  \sqrt{ 8\pi G_{N}}$, 
where $G_{N}$ is the gravitational constant.  If the gravitino is the LSP, the next-to-lightest 
supersymmetric particle (NLSP) decays to its standard model partner and a gravitino. 
The NLSP is a weak-scale particle decaying gravitationally and so has a 
natural lifetime of the order of: $M_{RPS}^{2}/M^{3}_{weak} \sim $ 10$^{4}$-10$^{8}$s ($\tau_{NLSP}$). 
This extraordinarily long lifetime renders the 
decay of the NLSP particle undetectable in a typical collider detector.

The superWIMP is a good candidate for dark matter and as such the gravitino LSP 
scenario  is constrained by measurements of non-baryonic cold dark matter density. 
The NLSP decays will  deposit electromagnetic \cite{NLSPEM} and 
hadronic \cite{NLSPHADRONIC} energy into  the universe, which may upset the successful predictions 
of standard big bang nucleosynthesis (BBN). Also, it is possible these NLSP 
decays may  distort the cosmic microwave  background (CMB) from its observed 
Planckian spectrum. In addition, photons produced in NLSP decays 
are subject to bounds on the diffuse photon flux. However, it turns out the superWIMP 
scenario can satisfy all these constraints.

Neutralino NLSPs are highly disfavored \cite{GRAVITINO8}
\cite{GRAVITINO12}. However, slepton and sneutrino NLSP scenarios 
have been analyzed and found to be safe \cite{GRAVITINO7}\cite{GRAVITINO8}
with the result that  the most natural NLSP candidates are sleptons, 
particularly the right-handed stau. Cosmological constraints exclude the 
upper range of $\tau_{NLSP}$ since, if the decays occur in a colder universe, then
the decay products would not be effectively thermalized. The CMB and BBN constraints, 
for typical thermal relic NLSP abundances, provide an upper bound on NLSP lifetimes 
that  excludes $\tau_{NLSP}$ above roughly a year.

Late NLSP decays may even resolve the leading BBN $^{7}$Li anomaly 
by reducing the predicted $^{7}$Li abundance to  the low 
values favoured by observations  \cite{GRAVITINO2}
  \cite{GRAVITINO3}. The NLSP lifetime that best
resolves the $^{7}$Li anomaly is 3$ \times$ 10$^{6}$ s  \cite{GRAVITINO5}, 
of the order of a month. Another motivation follows from considerations 
of leptogenesis \cite{LEPTOGENESIS}. In the gravitino LSP scenario, the gravitino does not decay.
Thus, the reheat temperature is bounded only by the overclosure constraint 
on the gravitino density. For  a gravitino mass $M_{\tilde{G}} \sim$ 100 GeV, reheat temperatures
as high as $\sim$ 10$^{10}$ GeV are allowed consistent 
with thermal leptogenesis \cite{REHEATTEMP1}\cite{REHEATTEMP2}.
Further connections between gravitino LSPs and  leptogenesis 
 are discussed in Ref. \cite{CONNECTIONS}.

The collider implications of a gravitino LSP with a charged slepton NLSP with lifetime 
under (but not much under) a year  have been investigated \cite{FENGSTER}.
In particular, the possibility of trapping sleptons in material (water tanks) placed just
outside Large Hadron Collider (LHC) or International Linear Collider (ILC) detectors 
was investigated. In this approach the  material in which the sleptons are
trapped is moved to an underground location (underground resevoirs) in order that 
slepton decays may be observed in a relatively background-reduced environment. 
At the Large Hadron Collider (LHC), the investigators found that, depending on the mass scale
of supersymmetry and the luminosity of the machine, tens to thousands of sleptons 
may be trapped each year. This implies that percent level studies of
sleptons  may be performed with resulting insights into
supergravity, supersymmetry breaking, dark matter and dark energy \cite{FENGSTER}. 

In this paper, we report a method of measuring the slepton liftime in a typical
LHC detector modelled on ATLAS,  using the decays of sleptons trapped in the detector itself. 
 No use is made of additional water traps or underground
resevoirs. The cosmic ray background is essentially eliminated by the depth of the
detector underground ($\sim$ 100 m) and by only utilizing upward going slepton
decays. The background from upward going neutrino
interactions, giving rise to upward going muons, is eliminated by considering 
only upward going decay products originating from inside a fiducial volume 
defined to be within  the outer RPC and TGC layers. 

Our analysis shows that definite evidence  of the existence 
of the decays of trapped sleptons would be available in the first year of running with
a received luminosity of  10 fb$^{-1}$. 
A measurement of the lifetime  to 4\% is possible after 3 years at nominal luminosity 
and running conditions and a received luminosity of 120 fb$^{-1}$. An observation
 of a trapped slepton signal in this manner would presumably justify the more 
 extensive and expensive water traps and instrumented 
resevoirs discussed elsewhere.

\section{Slepton Properties}

Current collider bounds from the null searches for long lived tracks at LEP-II \cite{BOUNDS}
require the slepton mass ($m_{\tilde{l}}$) to be greater than  99 GeV. Cosmology 
also has something to say about the slepton mass. In the gravitino superWIMP dark 
matter scenario, assuming that superWIMPS provide most, if not all, of the cold 
non-baryonic dark matter with density in the range $0.094 < \Omega_{DM}h^{2} < 0.124$,
the mass of the slepton is constrained to the range $m_{\tilde{l}} \sim$ 700-1000 GeV. 
Sleptons this heavy will be difficult to study at the LHC.

However, if gravitinos are  produced during reheating,  then for the  reheating temperatures 
preferred for leptogenesis as discussed above ( $\sim$ 10$^{9}$ GeV),  
gravitinos could form all of the non-baryonic dark matter for slepton masses as 
low as 120 GeV \cite{LOWSLEPTONMASS}. Alternatively, if the neutralino ($\tilde{\chi}$) is slightly heavier 
than the slepton, then  it will freeze out at the same time as the slepton and then decay 
to the slepton, thus combining its relic density with that of the slepton. This 
permits much lower slepton masses to  produce the correct gravitino dark matter density. 
For example, in minimal supergravity (mSUGRA)  with $A_{0}$ = 0, tan$\beta$ = 10, and $\mu > 0$, the required 
relic density may be achieved in the region $\tilde{\tau}_{LSP} \tilde{\chi}_{LSP}$ at 
$M_{1/2} = $ 300 GeV, where $M_{\tilde{\tau}}  \approx M_{\tilde{\chi}}
\approx $ 120 GeV \cite{BORDER}. When the gravitino is not the LSP, this is in the 
excluded Òstau LSPÓ region.

Assuming the lepton mass is negligible, the width of the slepton decay 
$\tilde{l} \rightarrow l\tilde{G}$ is:
\begin{equation}
\Gamma(\tilde{l} \rightarrow l\tilde{G}) \rightarrow \frac{1}{48\pi M_{RPS}^{2}}
\frac{M_{\tilde{l}}^{5}}{M_{\tilde{G}}^{2}}\left[1 - 
\frac{M_{\tilde{G}}^{2}}{M_{\tilde{l}}^{2}}\right]^{4}
\end{equation}
As the gravitino decays  gravitationally, the decay is determined only by the
lepton mass, the gravitino mass and the Planck mass - no SUSY  parameters enter. 

 In the  scenario  considered in this paper where the  gravitino is the  LSP, 
 the lower bound on $M_{1/2}$  (300 GeV)  is determined by the requirement of a stau
NLSP. The number of trapped staus generated rapidly diminishes as $M_{1/2}$ 
 increases and an  upper bound on $M_{1/2}$ ( = 800 Gev/c$^{2}$) can be defined to be
  where only a few staus  may be trapped in ATLAS per year. 
 %The superpartner spectra for various M1/2 in these models are given in Fig. 4.

Knowledge of the slepton range in matter is a key aspect of this analysis.  Charged
particles passing through matter lose energy by emitting radiation and by ionizing atoms.
In this analysis, we assume  the slepton loses energy in a similar way to a 
very heavy muon of the same mass, where ionization losses dominate over energy loss by
radiation. In this case, the GEANT3 simulation package \cite{GEANT3}
was used to estimate the range. Another author \cite{FENGSTER} 
directly estimated  the energy loss of the 
slepton per gcm$^{-2}$ using the Bethe Bloch formula.
The results of the GEANT simulation are in  excellent agreement 
with the direct calculation   for the range of a stau.
\FIGURE[htb]{\epsfig{file=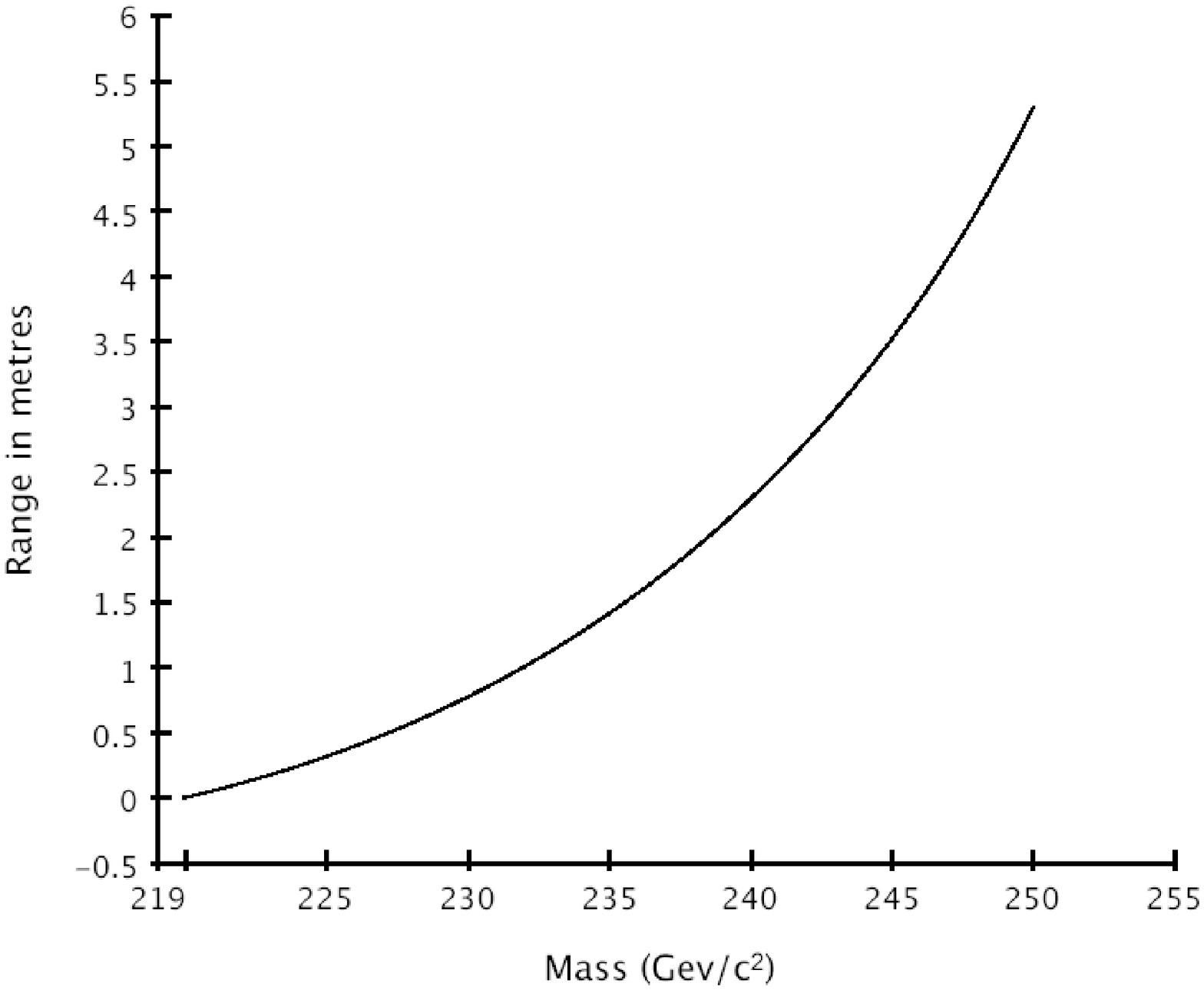,width=0.6\textwidth} 
        \caption{The range in lead of a stau with mass   
         219 GeV c$^{-2}$ as a function of energy, estimated using a GEANT3
	   simulation of the  stau range.}
	\label{fig:srange}}

\section{The Model of the Detector  and its Surroundings} \label{sc:model}

%To study trapped stau decay, the proton-proton collisions of the LHC are 
%simulated using HERWIG 6.507 \cite{Herwig:2002,Herwig2:2002} interfaced with 
%ISAJET 7.71 \cite{Isajet:2003,Isasusy:1993,Isawig:2002} in order to obtain the 
%SUSY mass spectrum. 

The ATLAS detector and experiment caverns, and the ATLAS 
rock overburden and underburden are all modelled in GEANT 3.21 
\cite{GEANT3}. Our detector model,  shown in Figure~\ref{atlas},  is based closely 
on the ATLAS detector \cite{ATLAS},  as described in the ATLAS Technical
Design Reports \cite{ATLASTDR}.  This model includes the beam pipe and shielding, 
the inner detector (the pixel detector, the semiconductor tracker (SCT) and the transition 
radiation tracker(TRT)), the calorimetry (the electromagnetic calorimeter (ECAL), the hadronic
tile calorimeter (HCAL), the liquid argon endcap calorimeter (LAr) and the 
forward calorimeter (FCAL)), the magnet system (the barrel solenoid and the 
barrel and endcap toroids) and the muon system (the monitored drift tube 
chambers (MDT), the cathode strip chambers (CSC), the resistive plate chambers
(RPC) and the thin gap chambers (TGC)). 

All  volumes are filled with a material that is a smeared average of the materials 
found in the corresponding actual detector volumes,  where these averages 
 correspond to the radiation lengths of each of the ATLAS components.
An ideal toroidal field is  used to model the magnetic field inside the barrel 
and endcap toroid models,  and the barrel solenoid has a uniform field of 2 T. 

%\fverb!\FIGURE[pos]{body}!\fverbdo
%\smallskip
%\FIGURE{\epsfig{file=atlas_3dbox_new.eps,width=0.4\textwidth} 
 %      \caption{The GEANT3 model of the ATLAS detector.}
 %      	\label{fig:3datlas}}

%\fverb!\FIGURE[htb]{body}!\fverbdo
%\smallskip
\FIGURE[htb]{\epsfig{file=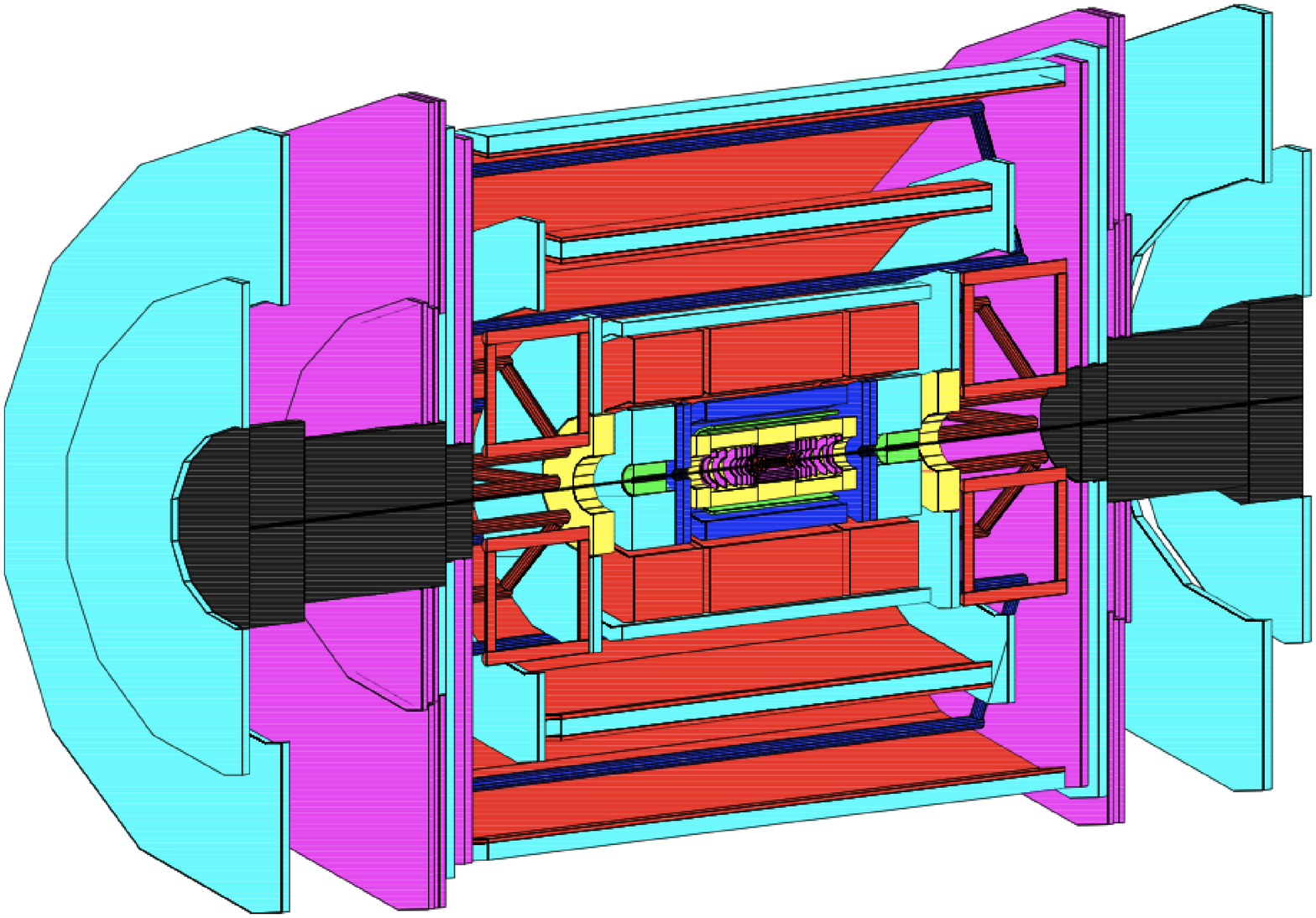,width=9cm, angle=-90} 
        \caption{GEANT model of an  LHC detector based closely on ATLAS.}
	\label{atlas}}

The muon system, particularly the RPCs and TGCs, has the most direct impact on 
this study of trapped stau decay. The RPCs and TGCs are dedicated trigger 
layers and, in combination with the precision of the MDTs, can allow the muon 
system to stand on its own as a muon detector. The RPCs lie in the barrel 
section of the detector and are separated into three layers. The central layer 
is referred to as the pivot layer, where hits in this layer are projected 
outward to the inner and outer RPC layers in coincidence windows (cones about
the direction of the incoming muon that are dependent on the muon's transverse
\footnote{The transverse direction is orthogonal to the direction of the beam.}
momentum) to search for matching hits. The outer layer is used only to search 
for high transverse momentum muons. The TGCs are split into five layers in each
of the ATLAS endcap sections. All but one of these layers is in the form of a
doublet, containing two carbon dioxide--\emph{n}-pentane gas volumes that are 
each layered between graphite cathodes and separated by paper honeycomb. The 
other TGC layer is a triplet, consisting of three gas volumes. The outermost 
doublets are the endcap pivot layers, where hits there are projected inward to 
the other TGC layers and matching hits are searched for in the same fashion as 
the RPCs.

%Figure \ref{fig:cavern} shows ATLAS sitting in UX15 surrounded by 
%the other experimental caverns and rock layers. 

%\begin{figure}
%\centering
%\includegraphics[width=\textwidth]{pictures/ch4/cavern1}
%\caption[The ATLAS cavern system]{The image shows the ATLAS model sitting in 
%         the main experimental hall, UX15, surrounded by the two counting
 %        rooms, US15 and USA15, and the four access shafts. The density in 
%         gcm$^{-3}$ and height in m of each rock overburden layer is also 
%         shown. The density of the rock underburden is 2.5 gcm$^{-3}$. The 
%         surface building, SX1, is also included.}
%\label{fig:cavern}
%\end{figure}

%\FIGURE{\epsfig{file=cavern1.eps,width=5cm} 
%        \caption{The image shows the ATLAS model sitting in 
%         the main experimental hall, UX15, surrounded by the two counting
%         rooms, US15 and USA15, and the four access shafts. The density in 
%         gcm$^{-3}$ and height in m of each rock overburden layer is also 
 %        shown. The density of the rock underburden is 2.5 gcm$^{-3}$. The 
%         surface building, SX1, is also included.}
%	\label{fig:cavern}}

\section{The Monte Carlo Simulation}

An R-parity conserving mSUGRA  SUSY breaking model is assumed, with the common scalar mass $m_{o}$ 
and the common soft breaking parameter $A_{o}$ taken to be 0 GeVc, the 
ratio of the Higgs vacuum expectation values tan$\beta$ is taken to be 10, the 
sign of the higgsino mass term  $\mu$ taken to be greater than 0, and the 
common gaugino mass $M_{1/2}$ varied between 300 GeV and 800 
GeV in increments of 100 GeV. This one-dimensional parameter 
set lies in a region of the mSUGRA parameter space that requires the gravitino 
to be the LSP, otherwise it would be an excluded stau 
LSP region \cite{FENGSTER}. The lower bound on $M_{1/2}$ was chosen to have a 
stau NLSP, and the upper bound is chosen so 
that the SUSY cross-section is high enough to have enough staus created each 
year to make this analysis useful. Table \ref{table:xsec} lists the SUSY 
cross-section that corresponds to each value $M_{1/2}$.

\begin{table}[htb]
\centering
\begin{tabular}[c]{ c c c c c c c  }
\hline \hline
$M_{1/2}$ (GeV) & 300 & 400 & 500 & 600 & 700 & 800  \\
\hline
$\sigma_{SUSY}$ (pb) & 20.2 & 4.87 & 1.51 & 0.542 & 0.219 & 0.096  \\
\hline
\end{tabular}
\caption[SUSY cross-sections]{The SUSY cross-sections that correspond to the
         values of $M_{1/2}$ considered.}
\label{table:xsec}
\end{table} 

To study trapped stau decay, the proton-proton collisions of the LHC are 
simulated using HERWIG 6.507 \cite{HERWIG} interfaced with 
ISAJET 7.71 \cite{ISAJET} in order to obtain the 
SUSY mass spectrum. We consider sparton, gaugino and/or slepton production generated using 
IPROC=3000 \cite{HERWIGMC} where  the three following processes are
implemented:
\begin{itemize} 
\item 2-parton $\rightarrow$ 2-sparton processes. All QCD sparton, i.e. squark and gluino, 
pair production processes are implemented;
\item 2-parton $\rightarrow$ 2-gaugino or gaugino+sparton processes. 
All gaugino, i.e. chargino and neutralino, pair production processes and 
gaugino-sparton associated production processes are implemented; and,
\item 2-parton $\rightarrow$ 2-slepton processes. All Drell-Yan slepton production 
processes are implemented.  
\end{itemize}

The stau lifetimes that we consider are 7, 30 and 90 days.
Each stau is given a random creation time throughout a year during 
those periods when the LHC beam is on, and a random decay time according to its 
lifetime spectrum.

After being assigned a lifetime, creation time and decay time, the staus are
passed through the GEANT ATLAS model. GEANT tracks the staus until the stau 
momentum falls below 0.001 GeV/c$^{2}$, at which point it is assumed the
stau has become trapped. It is also assumed that all staus will stop before
decaying \footnote{This is valid because the lifetimes considered here are much
greater than the time required for the stau to travel even 1 km away from ATLAS
through the rock overburden or underburden.}. Also, because the stau is heavy
and has an electromagnetic but not a colour charge, GEANT treats it like an 
extremely massive muon as it travels through matter, losing most of its energy 
through ionization effects.

%The model of the ATLAS overburden consists of eight layers of 
%rock of differing uniform densities, extending up to 78.91 m above the 
%interaction point. The underburden has two layers of rock of uniform density, 
%extending downward 1000.0 m below the interaction point. Each of these rock 
%layers extends out to $\pm$1000.0 m in x and z. Table \ref{table:overburden} 
%shows the y-position and thickness of each of the rock layers, as well as the 
%density of the rock that forms them. A mixture of 52.9\% oxygen, 33.7\% 
%silicon, 4.4\% calcium, 3.4\% aluminum, 1.6\% sodium, 1.4\% iron, 1.3\% 
%potassium and 1.0\% hydrogen, by mass, gives an approximate composition of 
%cement, which, with a varying density, then models the rock overburden and 
%underburden.

%\begin{table}[htb]
%\centering
%\begin{tabular}[c]{ c c c }
%\hline \hline
%Y$_{centre}$ & Thickness & $\rho$ \\
%(cm) & (cm) & (g cm$^{-3}$) \\
%\hline
%7741.0 & 300.0 & 2.4 \\
%7431.0 & 320.0 & 2.3 \\
%7081.0 & 380.0 & 2.5 \\
%6741.0 & 300.0 & 2.35 \\
%6266.0 & 650.0 & 2.4 \\
%4776.0 & 2330.0 & 2.4 \\
%3126.0 & 970.0 & 2.5 \\
%700.0 & 3882.0 & 2.45 \\
%-8120.5 & 6879.5 & 2.5 \\
%-15000.0 & 85000.0 & 2.5 \\
%\hline
%\end{tabular}
%\caption{There are a total of eight rock layers comprising the
%       overburden and two rock layers forming the underburden of ATLAS. Each 
%        of these extends to $\pm$1000.0 m in x and z. This table shows the 
%         y-position of the centre of each rock layer, as well as its thickness 
%        and the density of the rock in it. }
%        \label{table:overburden}
% \end{table}

Once the stau has stopped in ATLAS, it is decayed by GEANT, and the resulting muons and
pions down the decay chain are tracked as they pass through ATLAS. At that 
point, a model trigger decision is applied that mimics the decision made by the
cosmic ray muon trigger. Based on \cite{KOHNO}, it is assumed that the
trigger in the barrel region has an 83$\%$ efficiency, and the trigger in the
endcap region has a 95$\%$ efficiency.

\section{Triggering on Trapped Slepton Decay} \label{sc:trigger}

Once the LHC attains its design luminosity of 10$^{34}$ cm$^{-2}$s$^{-1}$,
ATLAS will become subject to more than 10$^{9}$ events per second. Because of
their high cross-sections, QCD processes will dominate these events and any
rare signals, like SUSY processes, will be swamped by this
background. However, ATLAS utilizes a three-tier trigger system to extract
these rare events and successively reduce the data rate to a final recordable 
rate of approximately 200 Hz.

The low-level trigger, level 1 (LVL1), is hardware-based and consists of custom
electronics operating at the LHC bunch crossing frequecy of 40 MHz with a 
latency time of just 2.5 $\mu$s. Up to 256 distinct trigger types that rely on
information from the calorimeters and muon detectors may be programmed into the
LVL1 trigger menus. The level 2 (LVL2) trigger and event filter (EF) make up 
the software-based high-level trigger that runs on dedicated processor farms.

The ATLAS cosmic ray trigger consists of a set of dedicated LVL2 cosmic ray 
muon trigger algorithms that rely heavily on the LVL1 muon trigger 
architecture. In the barrel region, the LVL1 cosmic ray muon trigger 
\cite{CONVENTIKAYANA} requires coincidences in both the $\eta$ and
$\phi$ directions of the inner two RPC layers. If the RPC pivot layer 
registers a hit, then the trigger searches for a coincidence in the inner RPC
layer across all sectors of the inner RPC layer that are electrically 
connected. This loosens the pointing requirement that is inherently built into
most LVL2 trigger algorithms, and corresponds approximately to the muon 
pointing to within 2 m of the interaction point. The timing of the cosmic rays
is determined using  the MDT chambers. 

The physical cross-section of the TGC endcaps is very small with respect to the
cosmic ray muons that arrive from above essentially parallel to the TGC
surface. Consequently, to boost the number of triggers that can occur, dummy 
hits are introduced into the large wheels to force a trigger to fire when only 
a single layer of the TGC triplet is actually hit. Since the layers of the TGC 
triplet are so closely spaced, any particle originating from the interaction
point will likely register a hit in all three layers, with a track that may be
traced back to the interaction point and that has a matching track in the inner
detector. Cosmic ray muons will generally not point to the interaction point,
so they are easily distinguished as cosmic rays.

%The LVL2 algorithms search the MDT chambers around the RPC and TGC layers in 
%the barrel and endcaps that were hit in order to extract precision information
%for the cosmic ray muon events. Because of their position measurement accuracy,
%the MDT chambers allow for precise timing of the cosmic ray events. 

According to ATLAS, then, the signal of the trapped staus will look very 
similar to that of cosmic ray muons. The stau decay products will not generally
originate from the interaction point, nor will they fit nicely into the bunch 
crossing periods of the LHC. As such, the ATLAS cosmic ray muon trigger is a 
natural choice to search for the signal of the trapped stau decay. Because the 
cosmic ray trigger can perform precision timing of the muons, it will also work
for decay products originating both from within and outside of ATLAS for  
staus trapped either within ATLAS or in the rock surrounding the main ATLAS 
experiment cavern.

\section{Backgrounds} \label{sc:background}

 In order to eliminate  the overwhelming background produced 
by the proton-proton collisions, as well as the background from beam-halo and 
beam-gas rates \cite{BOONE}, we propose to   
search for the signal of trapped stau decay only when the LHC
beam is off. The only remaining background sources to 
 the trapped stau signal arise from cosmic rays and upward-going 
neutrino-induced muons.  Neutrino-induced upward-going staus may also
be a possiblity \cite{ZHANG}. 

ATLAS is bombarded by cosmic ray muons at a rate of kHz, which would likely 
completely shadow the trapped stau signal. This large flux can be immediately 
dealt with by discarding all downward-going tracks, which is well within the 
capacity of the ATLAS trigger. While this would cut the number of signal events
roughly in half, the signal to background ratio will be significantly enhanced
by only considering  upward-going events. 

The background from upward-going muons created in interactions in the
rock beneath ATLAS  can also be eliminated. The ``ATLAS-like'' cosmic ray trigger
considered here  requires that the particle must pass through the inner two RPC 
layers and within  2m of the interaction point (in the barrel region), or it 
must pass through  the innermost TGC layer (in the endcap). In order to
rule out muons entering ATLAS from below, we restrict the trapping region 
for  the staus comprising our  signal to be inside the outer RPC layer (within 
a radius of 1027.5 cm)  and inside the inner TGC layer
(inside a z half-length of 1291 cm). We also require the
 particles to be  upward going. Looking only at particles trapped within
the above-defined fiducial volume, we can be sure that we are not seeing
anything but signal.

\section{Slepton Lifetime Determination} \label{sc:meth}

The ratio of the number of stau decays seen by the model trigger to the total 
number of staus created gives the geometric detector acceptance, $\alpha$. 
Table~\ref{table:acceptance} gives $\alpha$ 
for upward-going staus trapped within the fiducial volume
 as a function of both $M_{1/2}$ and 
lifetime, where a total of 10$^{6}$ staus were produced at each point and 
$\alpha$ is an average over six different trials.

%\begin{figure}[htb]
%\centering
%\includegraphics[width=0.7\textwidth]{muonacc_new.eps}
%\caption[ATLAS geometric acceptance]{The ATLAS geometric acceptance $\alpha$ is
%         shown as a function of $M_{1/2}$ and lifetime. The acceptance of both
%         upward- and downward-going particles is shown in relation to the total
%         acceptance. Note that the upward-going acceptance is greater than the
%         downward-going acceptance.}
%\label{fig:accpt}
%\end{figure}

\begin{table}
\centering
\begin{tabular}[c]{ c | rcl | rcl | rcl }
\hline \hline
$M_{1/2}$ (GeV) & \multicolumn{3}{c}{7 days} &
\multicolumn{3}{c}{30 days} & \multicolumn{3}{c}{90 days}\\
\hline
300 & 713.6 & $\pm$ & 14.5 & 709.5 & $\pm$ & 34.8 & 693.8 & $\pm$ & 21.5\\
400 & 673.1 & $\pm$ & 72.4 & 705.1 & $\pm$ & 23.0 & 718.9 & $\pm$ & 20.3\\
500 & 722.8 & $\pm$ & 29.0 & 706.3 & $\pm$ & 27.1 & 702.2 & $\pm$ & 15.2\\
600 & 713.8 & $\pm$ & 37.1 & 702.5 & $\pm$ & 27.8 & 685.5 & $\pm$ & 19.9\\
700 & 703.5 & $\pm$ & 32.0 & 708.8 & $\pm$ & 29.2 & 716.4 & $\pm$ & 8.2\\
800 & 706.3 & $\pm$ & 33.7 & 702.6 & $\pm$ & 20.3 & 693.0 & $\pm$ & 13.5\\
\hline
\end{tabular}
\caption[Acceptance]{Acceptance ($\times 10^{6}$) of the ATLAS-like detector,
                     defined as the total number of upward-going stau decay
                     products observed by the model trigger relative to the 
                     total number of Monte-Carlo staus produced, from staus 
                     that are trapped within the fiducial volume.}
\label{table:acceptance}
\end{table}

The number of candidate stau decays that ATLAS will be sensitive to during a 
period when the beam is off will be proportional to the geometric acceptance
$\alpha$ multiplied by the stau production rate. Thus, the number of candidate
stau decays ATLAS will 
detect during a shutdown period as a function of elapsed time well be:
\begin{equation}
N_{hits} = \alpha N^{f}_{8,j}e^{-t/\tau},
\label{eqn:hits1yr}
\end{equation}
where $\tau$ is the stau lifetime, $t$ is the
amount of time since the beam shut off and $N^{f}_{8,j}$ is 
the number of trapped staus at beam shut off (see the Appendix for the 
complete definition).
%by equation \ref{eqn:analytic}. 
To increase the size of the data set, a sum over multiple 
years can be taken, leading to
\begin{equation}
N_{hits} = \alpha\left(\sum_{j=1}^{n}N^{f}_{8,j}e^{-t/\tau}\right),
\label{eqn:hitsmyrs}
\end{equation}
where $n$ is the number of years considered.

This proportionality may be expressed as
\begin{equation}
N  =  Ae^{-Bt}   \label{eqn:fit1} 
\end{equation}
\begin{equation}
   A = \alpha\sum_{j=1}^{n}N^{f}_{8,j}    \label{eqn:fit2} 
 \end{equation}
 \begin{equation}
  B = \frac{1}{\tau}.    \label{eqn:fit3}
  \end{equation}
%\label{eqn:fit}
As such, it is possible to model-independently determine the lifetime of the 
stau using equation \ref{eqn:fit1} to fit the distribution of the total 
number of candidate stau decay products seen as a function of elapsed time 
during a beam shut off period. The SUSY cross-section is embedded in the value of 
$A$ (see equation \ref{eqn:fit2}), and may thus be determined, although the 
depedence of $A$ on $\alpha$ makes this model-dependent.

Two scenarios are considered in this analysis. The first assumes  the LHC
has run for only one year at low luminosity (an integrated luminosity of 10 
fb$^{-1}$). The second scenario assumes that three years worth of data have 
been collected, with two years of LHC running at low luminosity (as above) and 
one year of LHC running at high luminosity (an integrated luminosity of 100 
fb$^{-1}$). An analysis completed after only one year of LHC running could be 
used to demonstrate whether a decay signal may be seen. Combining three years
worth of data would allow for a more accurate determination of the lifetime of
the decaying particle, while still within a reasonable amount of time to see results.

The LHC operating  schedule is taken from \cite{BAILEY1}
\cite{BAILEY2}, which results in 200 days of beam running, segmented by 3 day 
beam shut off periods and  a large 144 day beam shut down period over the 
winter. We use this large shut down period\footnote{If the lifetime of the stau is not long compared to the 3 day shut
off periods, then those periods could be used to perform this analysis, as 
well. This possibility is not considered here.}, where the assumed lifetimes of 7,
30 and 90 days are shorter than the total amount of time the beam is off.

To determine the statistical spread in these values, this fit was performed
on 400 distinct data sets for each scenario, using lifetimes of 7, 30 and 90
days, over the entire range of $M_{1/2}$. We only attempted this fit when the expected
 number of events was greater than ten with the bin size set so that there were at
 least 4 events in the first bin. Figure~\ref{fig:timefit} shows, as an example, 
  a distribution of the number of  stau decays versus the elapsed time during
  the shutdown period, summing over  3 years of data (2 years at low luminosity 
  and 1 year  at high luminosity). The solid curve depicts the fit to the data that 
   gives the stau lifetime as  29.9 $\pm$ 1.5 days (model independent) and 
   the SUSY cross-section as  18.9 $\pm$ 1.3 pb. Figure~\ref{fig:timesfit} shows
   the distribution of lifetimes obtained for 400 experiments, under the second scenario, for a true stau lifetime of 
   30 days at $M_{1/2}$ = 300 GeV/c$^{2}$. In Figure~\ref{fig:numberevents}, we see
   the corresponding distribution of number of events for the 400 experiments, and in
   Figure~\ref{fig:crossection} the distribution of the SUSY cross-section values.

\DOUBLEFIGURE[htb]{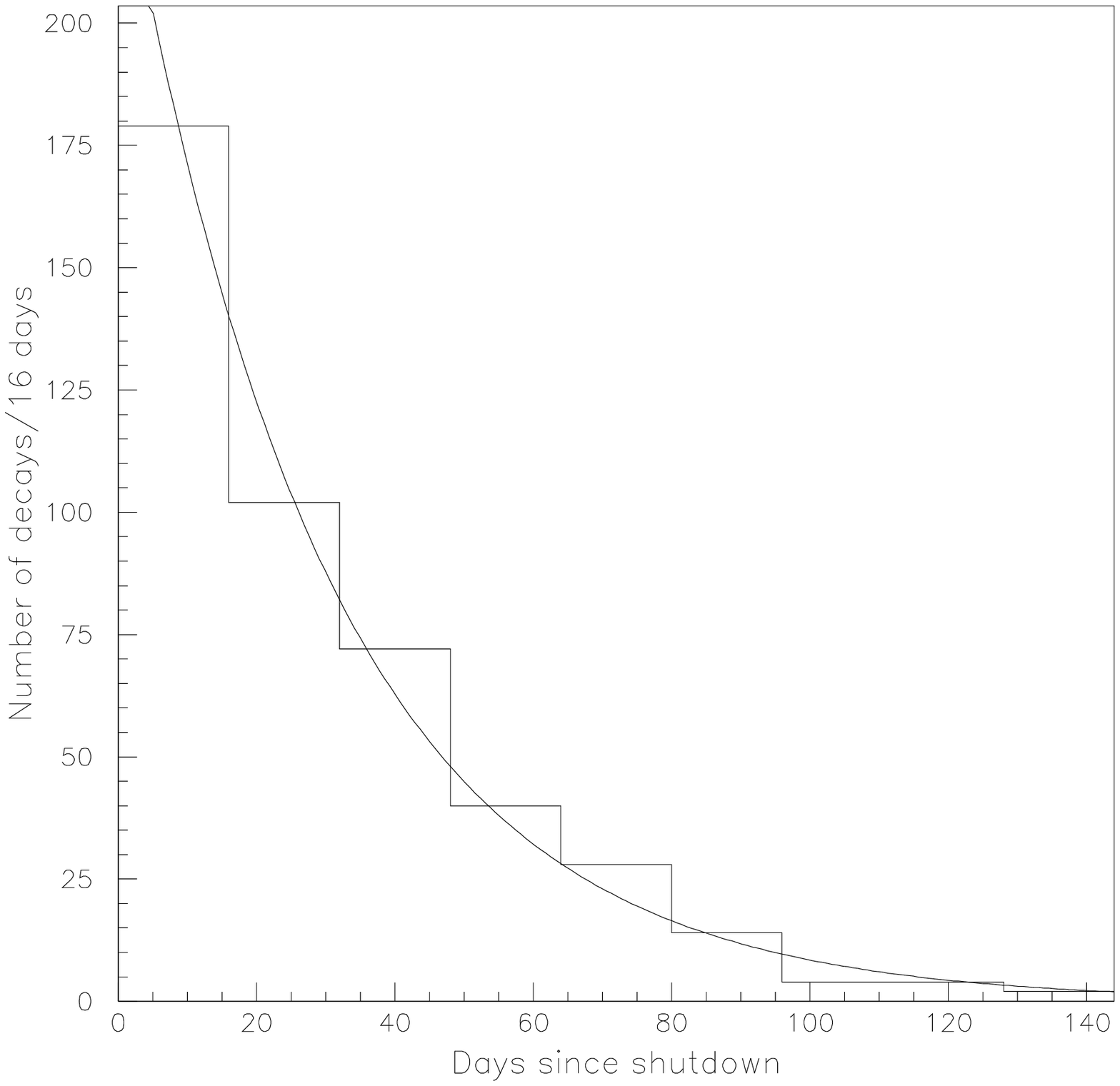, width=.4\textwidth}
{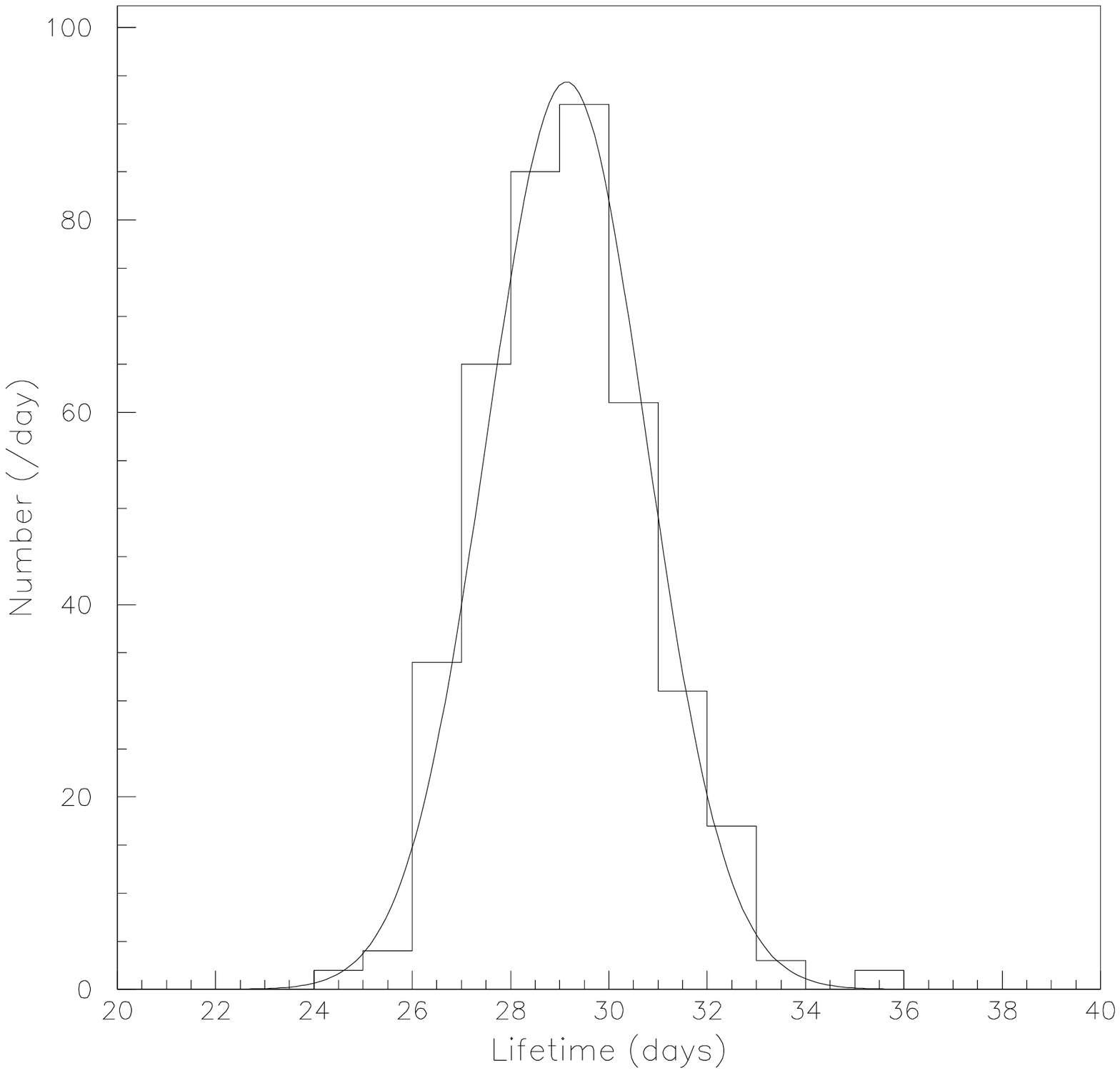, width=.4\textwidth}
{Sum over 3 years of data (2 years low luminosity and 1
                        year high luminosity) of the number of stau decay 
                        products seen by the model trigger as a function of 
                        elapsed time during the shutdown period. The fit of the
                        distribution results in a lifetime of 29.9 $\pm$ 1.5
                        days and a SUSY cross-section of 18.9 $\pm$ 1.3 pb. \label {fig:timefit} }
{The lifetime distribution for 400 experiments under
         the second  scenario where  the true stau lifetime is 30 days and the value of 
         $M_{1/2}$ is 300 GeV/c$^{2}$.\label{fig:timesfit}}
         
         \DOUBLEFIGURE[htb]{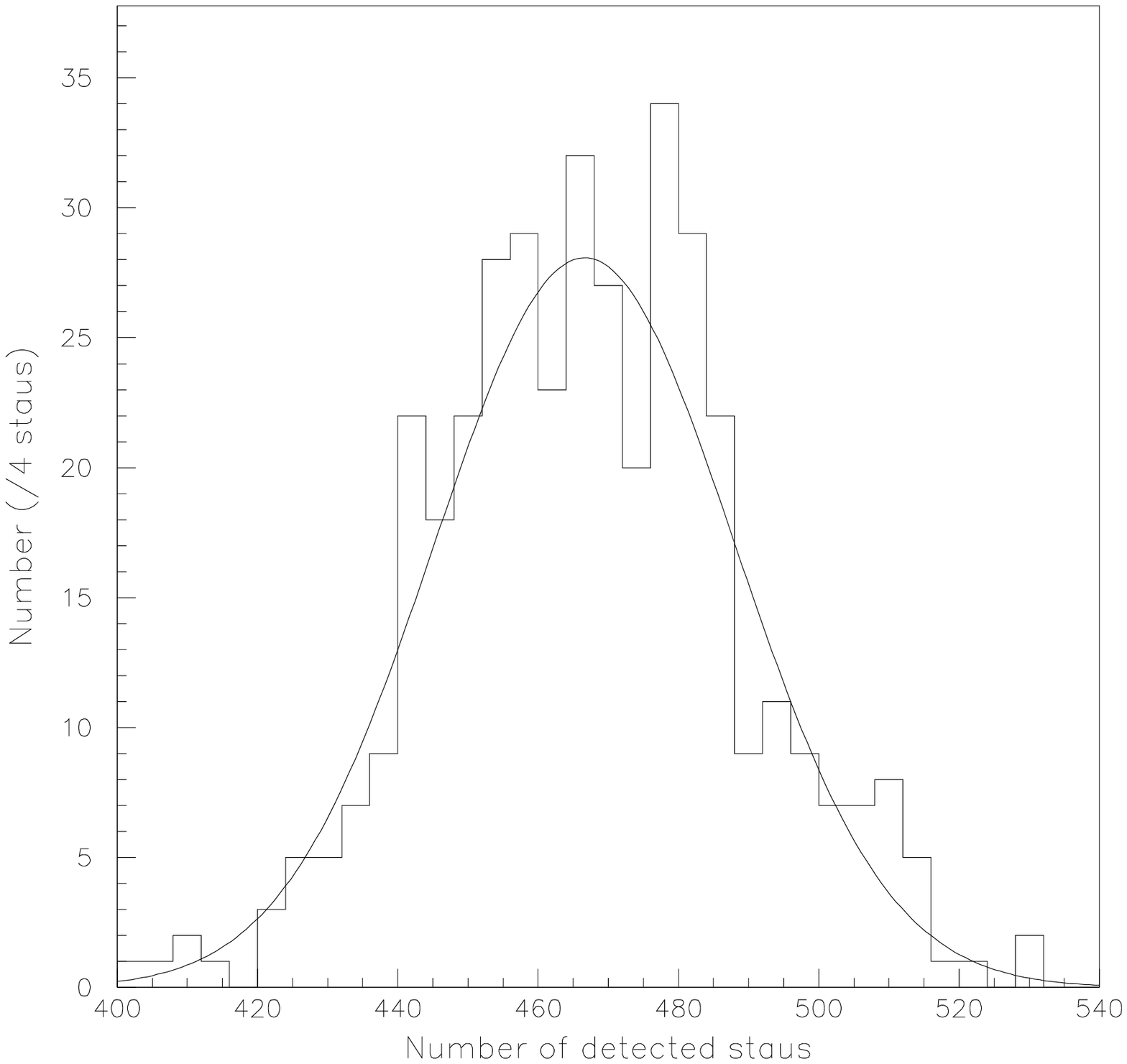, width=.4\textwidth}
{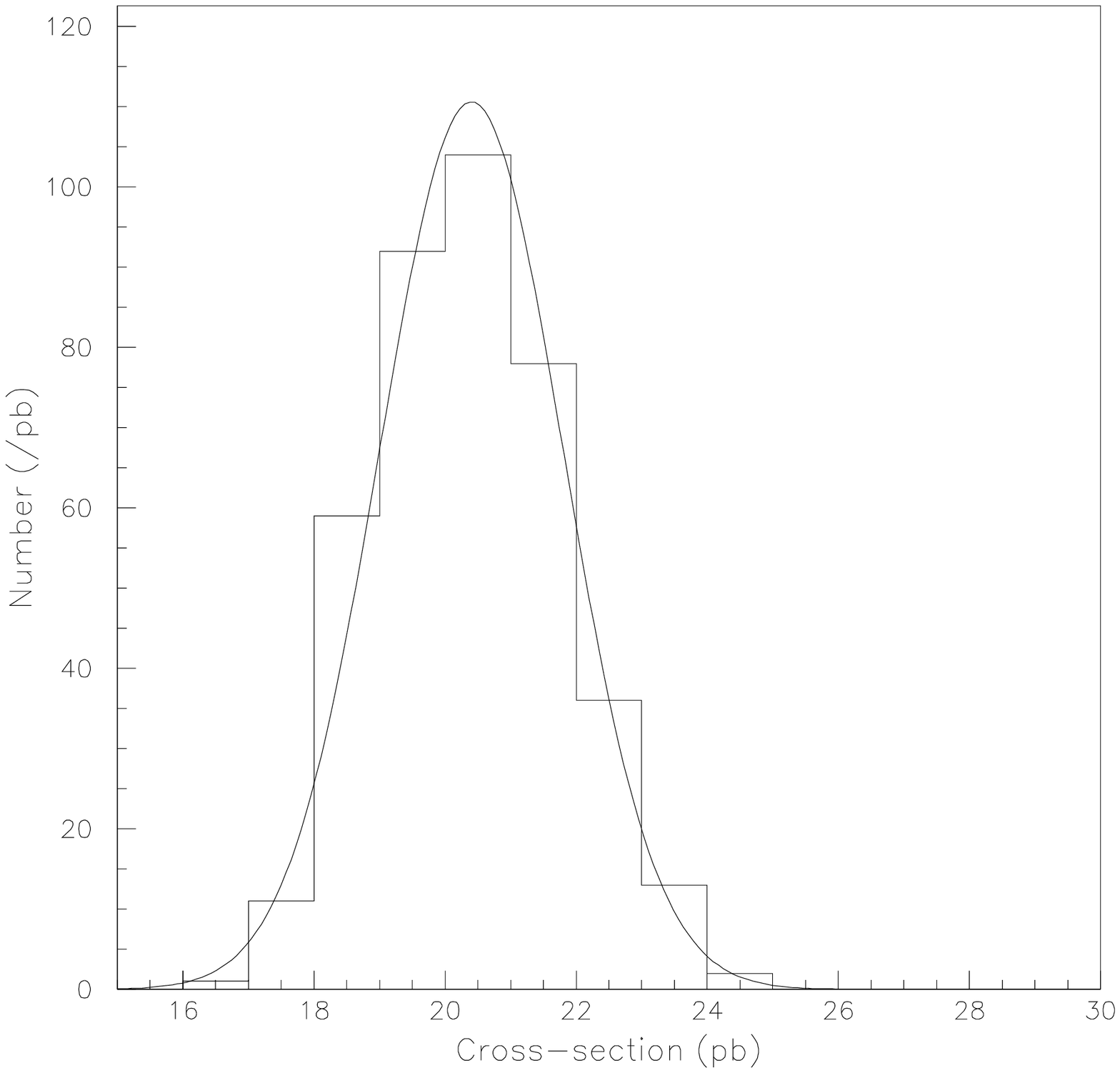, width=.4\textwidth}
{The distribution of the  number of events for 400 experiments under
the  second scenario  where  the true stau lifetime is 30 days and the value of 
         $M_{1/2}$ is 300 GeV/c$^{2}$.\label{fig:numberevents}}
{The SUSY cross-section  distribution for 400 experiments under
         the second scenario where  the true stau lifetime is 30 days and the value of 
         $M_{1/2}$ is 300 GeV/c$^{2}$.\label{fig:crossection}}

Tables \ref{table:spread7day}, \ref{table:spread30day}  and \ref{table:spread90day} give
the results for  the both scenarios where the slepton lifetimes are 7 days, 30 days and 90 days, 
respectively. One can see  for all scenarios  that, as the value 
of $M_{1/2}$ increases, the SUSY cross-section decreases and, consequently, so
does the accuracy of the stau lifetime and SUSY cross-section determination using trapped
staus.  

For  $\tau$ = 30 days, near  the cosmologically preferred value
 of the stau lifetime (37 days),  our results show (see Table~\ref{table:spread30day} ) that under Scenario 1 
 for $M_{1/2} \sim$ 300-400 GeV  and  SUSY  
 cross-section $\sigma_{SUSY}$=20.2-4.87 pb the stau lifetime and $\sigma_{SUSY}$ 
 determined agree with the actual values within a 
  fractional  (statistical) error on the measured value  of  $\sim \pm($25\%-60\%).
 Under this scenario we obtain  $\sim$ 40-10 signal events  after 1 year running at low luminosity. 
 For scenario 2, for  $  M_{1/2} \sim$ 300-600 GeV with  $\sigma_{SUSY}$=20.2-0.54 pb
 we can obtain a lifetime and cross-section estimate that agrees with the set values within a 
 fractional statistical uncertainty on the measured value of of  $\sim \pm$(5\%-60\%) and $\sim \pm$(7\%-70\%), 
 respectively,  with $\sim$ 467-13  signal events detected. 

Under Scenario 1,  for  a relatively short stau lifetimes of 7 days, we only have sensitivity for
$  M_{1/2} \sim$ 300 GeV with  cross-section $\sigma_{SUSY}$=20.2 pb. 
In this case,  the stau lifetime and SUSY cross-section  determined agrees with the actual value to within
the  fractional   (statistical) uncertainty  of  $\sim \pm$50\%  and 
  $\sim \pm$80\%, respectively,  on the measured value,  with $\sim$ 10 signal events detected. 
   We can do better under  scenario 2 where, for $  M_{1/2} \sim$ 300-500 GeV 
   and $\sigma_{SUSY}$ = 20.2-1.51 pb,  we can determine the stau lifetime and 
   $\sigma_{SUSY}$  to within the fractional   (statistical) uncertainty  on the measured value 
 of  $\sim \pm$(12\%-65\%).   In this case the number of detected events varies between $\sim$ 120 and $\sim$ 10.
 
 The longest  stau lifetime considered is 90 days. For Scenario 1 we have sensitivity for 
 $M_{1/2} \sim$ 300-400 GeV/c$^{2}$ and   $\sigma_{SUSY}$ = 20.2-4.87 pb. 
 Under this scenario, we  can determine the stau lifetime  $\sigma_{SUSY}$ 
 to within the  fractional   (statistical) uncertainty  on the measured value of  $\sim \pm$(23\%-30\%) 
  and  $\sim \pm$(13\%-24\%), respectively, with the number of recorded signal events varying 
  between  $\sim$ 83 and $\sim$ 21. For scenario 2, we can measure the stau lifetime over the range
   $M_{1/2} \sim$ 300-700 GeV/c$^{2}$ and   $\sigma_{SUSY}$ = 20.2-0.22 pb. 
   In this case, we can pin down  the stau  liftime and  $\sigma_{SUSY}$ to within  the fractional   
   (statistical) uncertainty  on the measured value 
 of  $\sim \pm$(7\%-40\%) and  $\sim \pm$(4\%-48\%), respectively, 
 with  the number of detected events varying between $\sim$ 990 and $\sim$ 12.

\begin{table}
\centering
\begin{tabular}[c]{c | c | rcl | rcl | c}
\hline \hline
$M_{1/2}$ (GeV) & Scenario & \multicolumn{3}{c}{1 year low} &
\multicolumn{3}{c}{2 years low 1 high} & Expected\\
\hline
 & $\tau$ & 6.8 & $\pm$ & 3.5 & 6.8 & $\pm$ & 0.8 & 7\\
300 & $\sigma_{SUSY}$ & 9.5 & $\pm$ & 7.3 & 20.9 & $\pm$ & 3.2 & 20.2\\
 & $N_{hits}$ & 10.2 & $\pm$ & 2.8 & 120.8 & $\pm$ & 10.5 & 10.0/119.9\\
\hline
 & $\tau$ & & $-$ & & 7.3 & $\pm$ & 2.7 & 7 \\
400 & $\sigma_{SUSY}$ &  & $-$ &  & 4.52 & $\pm$ & 1.75 & 4.87 \\
 & $N_{hits}$ & 2.4 & $\pm$ & 1.6 & 27.7 & $\pm$ & 4.5 & 2.3/27.3 \\
\hline
 & $\tau$ &  & $-$ & & 9.62 & $\pm$ & 6.31 & 7 \\
500 & $\sigma_{SUSY}$ &  & $-$ &  & 0.73 & $\pm$ & 0.48 & 1.51 \\
 & $N_{hits}$ &  & $-$ &  & 9.7 & $\pm$ & 2.8 & 0.8/9.1 \\
\hline
 & $\tau$ &  & $-$ &  &  & $-$ & & 7 \\
600 & $\sigma_{SUSY}$ &  & $-$ &  &  & $-$ & & 0.54 \\
 & $N_{hits}$ &  & $-$ &  & 3.4 & $\pm$ & 1.8 & 0.3/3.2 \\
\hline
& $\tau$ & & $-$ & & & $-$ & & 7 \\
700 & $\sigma_{SUSY}$ & & $-$ &  & & $-$ & & 0.22 \\
 & $N_{hits}$ &  & $-$ &  & 1.3 & $\pm$ & 1.6 & 0.1/1.3 \\
\hline
& $\tau$ & & $-$ & & & $-$ & & 7 \\
800 & $\sigma_{SUSY}$ & & $-$ &  &  & $-$ & & 0.096 \\
 & $N_{hits}$ &  & $-$ &  &  & $-$ & & 0.1/0.6 \\
\hline
\end{tabular}
\caption[Distributions for 7 day lifetime]{Fitted  and expected  values of stau lifetime, SUSY 
         cross-section and number of  events for a stau lifetime of 7
         days at both luminosity scenarios. The unit of lifetime is days
         and that of cross-section is picobarns.}
\label{table:spread7day}
\end{table}

% 30 days

\begin{table}
\centering
\begin{tabular}[c]{c | c | rcl | rcl | c}
\hline \hline
$M_{1/2}$ (GeV) & Scenario & \multicolumn{3}{c}{1 year low} &
\multicolumn{3}{c}{2 years low 1 high} & Expected\\
\hline
 & $\tau$ & 29.3 & $\pm$ & 7.3 & 29.1 & $\pm$ & 1.6 & 30\\
300 & $\sigma_{SUSY}$ & 19.9 & $\pm$ & 5.1 & 20.4 & $\pm$ & 1.4 & 20.2\\
 & $N_{hits}$ & 39.7 & $\pm$ & 6.2 & 466.6 & $\pm$ & 21.5 & 39.7/476.6 \\
\hline
 & $\tau$ & 33.9 & $\pm$ & 19.9 & 29.1 & $\pm$ & 3.6 & 30 \\
400 & $\sigma_{SUSY}$ & 2.83 & $\pm$ & 1.71 & 4.97 & $\pm$ & 0.72 & 4.87 \\
 & $N_{hits}$ & 10.1 & $\pm$ & 2.8 & 112.6 & $\pm$ & 10.4 & 9.5/114.2 \\
\hline
 & $\tau$ & & $-$ & & 31.6 & $\pm$ & 8.1 & 30 \\
500 & $\sigma_{SUSY}$ &  & $-$ &  & 1.37 & $\pm$ & 0.54 & 1.51 \\
 & $N_{hits}$ & 2.9 & $\pm$ & 2.1 & 34.8 & $\pm$ & 5.7 & 3.0/35.5 \\
\hline
 & $\tau$ &  & $-$ &  & 35.2 & $\pm$ & 21.2 & 30 \\
600 & $\sigma_{SUSY}$ &  & $-$ &  & 0.36 & $\pm$ & 0.26 & 0.54 \\
 & $N_{hits}$ &  & $-$ &  & 13.0 & $\pm$ & 3.5 & 1.1/12.6 \\
\hline
& $\tau$ & & $-$ & & & $-$ & & 30 \\
700 & $\sigma_{SUSY}$ & & $-$ &  & & $-$ & & 0.22 \\
 & $N_{hits}$ &  & $-$ &  & 5.6 & $\pm$ & 2.1 & 0.4/5.2 \\
\hline
& $\tau$ & & $-$ & & & $-$ & & 30 \\
800 & $\sigma_{SUSY}$ & & $-$ &  &  & $-$ & & 0.096 \\
 & $N_{hits}$ &  & $-$ &  & 2.8 & $\pm$ & 2.3 & 0.2/2.2 \\
\hline
\end{tabular}
\caption[Distributions for 30 day lifetime]{Fitted  and expected  values of stau lifetime, SUSY 
         cross-section and number of  events for a stau lifetime of 30
         days at both luminosity scenarios. The unit of lifetime is days
         and that of cross-section is picobarns}
\label{table:spread30day}
\end{table}

% 90 days

\begin{table}
\centering
\begin{tabular}[c]{c | c | rcl | rcl | c}
\hline \hline
$M_{1/2}$ (GeV) & Scenario & \multicolumn{3}{c}{1 year low} &
\multicolumn{3}{c}{2 years low 1 high} & Expected\\
\hline
 & $\tau$ & 87.5 & $\pm$ & 20.4 & 89.0 & $\pm$ & 6.0 & 90\\
300 & $\sigma_{SUSY}$ & 20.2 & $\pm$ & 2.6 & 19.9 & $\pm$ & 0.8 & 20.2\\
 & $N_{hits}$ & 82.8 & $\pm$ & 9.1 & 989.9 & $\pm$ & 30.9 & 83.7/1007.9 \\
\hline
 & $\tau$ & 65.4 & $\pm$ & 19.7 & 87.6 & $\pm$ & 10.7 & 90 \\
400 & $\sigma_{SUSY}$ & 5.06 & $\pm$ & 1.20 & 4.82 & $\pm$ & 0.40 & 4.87 \\
 & $N_{hits}$ & 20.8 & $\pm$ & 4.4 & 247.9 & $\pm$ & 16.3 & 20.9/251.8 \\
\hline
 & $\tau$ & & $-$ & & 83.8 & $\pm$ & 18.9 & 90 \\
500 & $\sigma_{SUSY}$ &  & $-$ &  & 1.53 & $\pm$ & 0.23 & 1.51 \\
 & $N_{hits}$ & 6.7 & $\pm$ & 2.6 & 75.3 & $\pm$ & 8.2 & 6.3/76.3 \\
\hline
 & $\tau$ &  & $-$ &  & 66.3 & $\pm$ & 18.2 & 90 \\
600 & $\sigma_{SUSY}$ &  & $-$ &  & 0.59 & $\pm$ & 0.13 & 0.54 \\
 & $N_{hits}$ & 2.4 & $\pm$ & 1.5 & 26.6 & $\pm$ & 5.3 & 2.2/26.6 \\
\hline
& $\tau$ & & $-$ & & 68.8 & $\pm$ & 27.8 & 90 \\
700 & $\sigma_{SUSY}$ & & $-$ &  & 0.23 & $\pm$ & 0.11 & 0.22 \\
 & $N_{hits}$ &  & $-$ &  & 11.5 & $\pm$ & 3.3 & 0.9/11.3 \\
\hline
& $\tau$ & & $-$ & & & $-$ & & 90 \\
800 & $\sigma_{SUSY}$ & & $-$ &  &  & $-$ & & 0.096 \\
 & $N_{hits}$ &  & $-$ &  & 5.4 & $\pm$ & 3.4 & 0.4/4.8 \\
\hline
\end{tabular}
\caption[Distributions for 90 day lifetime]{Fitted  and expected  values of stau lifetime, SUSY 
         cross-section and number of  events for a stau lifetime of 90
         days at both luminosity scenarios. The unit of lifetime is days
         and that of cross-section is picobarns.}
\label{table:spread90day}
\end{table}

\section{Conclusion} \label{sc:concl}

The sensitivity of the ATLAS-like detector to the decay of long-lived staus 
that become trapped in the detector was studied using a 
simplified model of the ATLAS detector. To completely eliminate the background from
cosmic rays and upward going neutrino-induced muons, only 
upward-going decay products  originating from inside a fiducial volume 
defined to be within  the outer RPC and TGC layers  were studied. In addition,
decays were only recorded during beam-off periods to further reduce backgrounds
from actual collisions.

 By fitting the  number of candidate stau decay products detected by a  model 
 cosmic ray muon trigger  during periods when the LHC beam is off 
  as a function   of the elapsed time during the beam shut off period, 
the model-independent stau lifetime and the  model-dependent SUSY cross-section 
can be determined.

 The fitting procedure  returns the stau lifetime and SUSY cross-section with up to approximately  
 ~5\% statistical accuracy on the measured value. For almost all situations considered 
 that allow a fit - those with $\sim$ 10 detected events or more  - the measured values of the
 slepton liftime and SUSY cross-section  lie
 within $\pm$1$\sigma$  of the actual value, where $\sigma$ is the statistical error on the measured value.
 We  found that for the two running scenarios considered, our analysis procedure has a good sensitivity to 
slepton lifetimes  of  $\sim$ 30 days  and a SUSY cross-section  greater than approximately 0.50 pb. 
It is interesting to note that  stau  lifetimes ($\sim$ 30 days  are cosmologically  favoured by models 
propounding a gravitino   LSP scenario.

%\section{Appendix 1}
%\section{Appendix 1} \label{app:math}
\appendix
\appendixpage
\addappheadtotoc

If we set the luminosity for year $j$ of running at the LHC to be $L_{j}$, then the 
stau production rate at ATLAS, assuming R-parity conservation, will be
\begin{equation}
R_{j} = 2\frac{L_{j}\sigma_{SUSY}}{T_{on}}\textrm{,}
\label{eqn:staurate}
\end{equation}
where $T_{on}$ is the total time over a year that the beam is on. The schedule
of beam operation assumed is given in \cite{BAILEY1}\cite{BAILEY2}, which 
gives a total of 200 days of LHC beam operation in one year. 
The integrated luminosity of the LHC is assumed to be 10 fb$^{-1}$ and 100 fb$^{-1}$ for
low and high luminosity years, respectively. 

Now, letting the number of undecayed staus remaining in or around ATLAS at 
beam turn-on be $N^{o}$ and the number remaining at beam turn-off be $N^{f}$,
the number of staus remaining as a function of time $t$ becomes
\begin{eqnarray}
N_{off} & = & N^{f}e^{-t/\tau} \nonumber \\
N_{on} & = & N^{o}e^{-t/\tau}+R\tau(1-e^{-t/\tau})\textrm{,}
\label{eqn:nonoff}
\end{eqnarray}
where $\tau$ is the stau lifetime.

Using equation \ref{eqn:nonoff} and labelling the eight LHC operating periods
with the index $i$ gives
\begin{eqnarray}
N^{o}_{1,1}&=&0 \nonumber \\
N^{o}_{1,j}&=&N^{f}_{8,j-1}e^{-t_{sd}/\tau}\textrm{,  } j \geq 2 \nonumber \\
N^{o}_{i,j}&=&N^{f}_{i-1,j}e^{-t_{off}/\tau}\textrm{,  }i\in[2,8]\nonumber \\
N^{f}_{i,j}&=&N^{o}_{i,j}e^{-t_{on}/\tau}+R_{j}\tau(1-e^{-t_{on}/\tau})
\textrm{,}
\label{eqn:iterative}
\end{eqnarray}
where $t_{sd}$, $t_{off}$ and $t_{on}$ are the amounts of time during a 
shutdown, offline and operational period of the LHC \footnote{The shutdown
period is a long 144 day period when the beam is shut off and the offline
periods are short three day down times between beam operational periods.}. This
set of equations can be used to determine the number of undecayed staus 
remaining at the beginning and end of each operational period, resulting in
\begin{eqnarray}
N^{f}_{i,j} & = & N^{o}_{1,j}e^{-i(t_{off}+t_{on})/\tau}e^{t_{off}/\tau}+
             \nonumber \\
 & & R_{j}\tau(1-e^{-t_{on}/\tau})\sum_{n=0}^{i-1}e^{-n(t_{off}+t_{on})/\tau}
             \nonumber \\
N^{o}_{1,j} & = & \sum_{m=1}^{j-1}R_{m}\tau(1-e^{-t_{on}/\tau})\sum_{n=0}^{7}
     e^{-n(t_{off}+t_{on})/\tau}\cdot \nonumber \\
 & & e^{-(j-1-m)(7t_{off}+8t_{on})/\tau}e^{-(j-m)t_{sd}/\tau}\textrm{.}
\label{eqn:analytic}
\end{eqnarray}
Meanwhile, the number of staus that decay during shutdown period $i$ of year $j$
is
\begin{eqnarray}
N^{decays}_{i,j} & = & N^{f}_{i,j}(1-e^{-t_{off}/\tau})\textrm{,  }i\in[1,7]
                    \nonumber \\
N^{decays}_{8,j} & = & N^{f}_{8,j}(1-e^{-t_{sd}/\tau})\textrm{.}
\label{eqn:ndecay}
\end{eqnarray}
Thus, the number of decays ATLAS will see will be the the total number of decays
that occur during a period when the beam is off multiplied by the geometric 
detector acceptance $\alpha$.

%========================================================================

\end{document}